\documentclass[a4paper,12pt]{article}
\usepackage{epsf,amsmath}
\usepackage{indentfirst}
\usepackage{amssymb}
\usepackage{hhline}
\usepackage{array,tabularx}
\def\@biblabel#1{#1.}
\makeatother

\begin{document}

\title{\Large\bf Merging of Components in Close Binaries: Type Ia Supernovae,
Massive White Dwarfs, and Ap stars}

\author{A. I. Bogomazov$^{1}$, A. V. Tutukov$^{2}$ \\
\small\it $^1$ Sternberg Astronomical Institute, Moscow State
University, \\ \small\it Universitetski pr. 13, Moscow, 119992, Russia, \\
\small\it $^2$ Institute of Astronomy, Russian Academy of
Sciences, \\
\small\it ul. Pyatnitskaya 48, Moscow, 109017, Russia \\
\small Astronomy Reports, volume 53, no. 3, pp. 214-222 (2009)}

\date{\begin{minipage}{15.5cm} \small
The ``Scenario Machine'' (a computer code designed for studies of
the evolution of close binaries) was used to carry out a
population synthesis for a wide range of merging astrophysical
objects: main-sequence stars with main-sequence stars; white
dwarfs with white dwarfs, neutron stars, and black holes; neutron
stars with neutron stars and black holes; and black holes with
black holes.We calculate the rates of such events, and plot the
mass distributions for merging white dwarfs and main-sequence
stars. It is shown that Type Ia supernovae can be used as standard
candles only after approximately one billion years of evolution of
galaxies. In the course of this evolution, the average energy of
Type Ia supernovae should decrease by roughly 10\%; the maximum
and minimum energies of Type Ia supernovae may differ by no less
than by a factor of 1.5. This circumstance should be taken into
account in estimations of parameters of acceleration of the
Universe. According to theoretical estimates, the most massive --
as a rule, magnetic -- white dwarfs probably originate from
mergers of white dwarfs of lower mass. At least some magnetic Ap
and Bp stars may form in mergers of low-mass main-sequence stars
($M\lesssim 1.5M_{\odot}$) with convective envelopes.
\end{minipage} } \maketitle \rm

\section{Introduction}

Here, we consider the role of component mergers in various close
binaries in the formation of some astrophysical objects and the
generation of one of the greatest natural phenomena -- Type Ia
supernovae (SN Ia).

Currently, the most popular explanation for SN Ia is mergers of
two carbon-oxygen white dwarfs (CO WDs) under the action of
gravitational radiation, provided the total mass of the merging
dwarfs exceeds the Chandrasekhar limit. This scenario for the
formation of SN Ia was suggested in the early 1980s (see, for
example, \cite{tutukov1981a,iben1984a}). Up to 40\% suggesting
that their progenitors may form an inhomogeneous group of objects
\cite{li2001a}. Apart from mergers of two CO dwarfs, possible
origins include the thermonuclear explosion of a WD in a
semi-detached system during the accretion of matter from its
companion, when the dwarf's mass exceeds the Chandrasekhar limit
\cite{whelan1973a}, and a thermonuclear explosion of a white dwarf
with a helium donor star companion; under certain conditions, the
mass of the WD may be lower than the Chandrasekhar limit
\cite{livne1990a}. Here, however, we consider the basic scenario
for the formation of SN Ia to be the merging of two CO dwarfs
whose total mass exceeds the Chandrasekhar limit. The high age
dispersion of SN Ia precursors, $\sim 10^8-10^{10}$ years
\cite{aubourg2007a} supports this scenario. The conditions for an
SN Ia explosion have also been considered, for example, in
\cite{yoon2007a}.

At the end of the twentieth century, observations of distant SN Ia
resulted in the discovery that the expansion of the Universe is
accelerating \cite{riess1998a,perlmutter1999a}. It was suggested
that SN Ia can be used as standard candles, and that their
magnitudes at the maximum brightness can be determined from the
slope of the light curve, according to the regular trends revealed
by Pskovsky \cite{pskovskii1984a}. Taking into account the most
popular SN Ia model, we suggest that the use of this type of
supernova as standard candles may encounter an obvious problem:
the total mass of the merging WDs may only slightly exceed the
Chandrasekhar limit, or be only a little lower than two
Chandrasekhar limits. In this case, the scatter of SN Ia maximum
luminosities could reach a factor of two. Average total mass of
the merging dwarfs decreases with time imitating accelerated
expansion of the Universe. Additional uncertainty is introduced by
the fact that current concepts do not exclude the possibility of a
thermonuclear explosion of two merged degenerate dwarfs whose
total mass is somewhat below the Chandrasekhar limit.

Optical observations of SN Ia revealed some inhomogeneity in their
properties. A subtype with lower luminosities, and probably lower
energy releases, can be distinguished (for example,
\cite{taubenberger2008a}). It remains unclear if this subtype is
associated with a decreased abundance of heavy elements or a lower
mass of the pre-supernova. SN Ia may also differ somewhat in the
expansion velocities of their envelopes \cite{pignata2008a};
again, it is unclear whether this is due to different explosion
energies or, for example, different orientations of the supernova
envelope relative to the observer \cite{roepke2006a}. The increase
in the frequency of SN Ia explosions at early stages in the
evolution of galaxies \cite{blanc2008a}, when the star formation
rate was higher than its present value \cite{shustov1997a},
provides evidence that relatively short-lived objects with
lifetimes of about $1-2\cdot 10^9$ years are also among SN Ia
progenitors. However, in most cases, SN Ia progenitors have longer
lifetimes \cite{forster2008a}. Some SN Ia are embedded into
circumstellar envelopes that are expanding with velocities of up
to 50 km/s \cite{patat2008a}, suggesting that these supernovae
evolved from symbiotic binaries with red-giant donors. This model
was suggested, for example, in \cite{tutukov1972a} to explain
novae outbursts, and in \cite{whelan1973a} to explain Type I
supernovae, which at that time were not yet divided into the Ia,
Ib, and Ic subtypes. A scenario in which degenerate components of
close binaries merge under the action of gravitational waves is
also possible, though unlikely, in this case. The observed
envelopes are then the remnants of the common envelopes, which
``dissolve'' in the interstellar medium on time scales of $\sim
10^5$ years.

In order to investigate the ``standardness'' of SN Ia, we have
calculated the rates of mergers of degenerate dwarfs with various
chemical compositions, constructed the mass functions of merging
WDs, and derived the average mass of merging dwarfs as a function
of the time since the onset of star formation in a galaxy under
various assumptions concerning the star-formation history. We also
consider the results of merging two WDs with a total mass below
the Chandrasekhar limit and of degenerate dwarfs with various
masses and ages.

The magnetic fields of non-interacting magnetic white dwarfs
(MWDs) lie in the interval 0.1-1000 MG, with the peak in the
distribution around 16 MG (see the review
\cite{wickramasinghe2000a}). It is commonly believed that MWDs are
the remnants of magnetic Ap and Bp stars. There are also some
indications that the MWD mass function is bimodal, with a second
maximum close to the Chandrasekhar limit. This maximum could
correspond to mergers of degenerate dwarfs. The rotational
velocities MWDs are also bimodal: they include both rapidly
rotating stars with rotational periods from roughly 700 s to
several hours, as well as virtually non-rotating MWDs, for which
formal estimates of their rotational periods yield values of the
order of 100 years. The rapidly rotating MWDs are naturally
through to arise from mergers of WD components in close binaries.
In addition, their rotation may have been substantially
accelerated by accretion during their evolution in a close binary.
Slow MWDs have most likely evolved from stars (either single, or
components of wide binaries) whose cores were strongly magnetized
and lost their angular momentum during interaction with the
extended envelope of the star. It can also not be excluded that
the magnetic field strongly affects the initial ``mass-radius''
relation, and thus the consequences of the evolution of close
binaries, such as the SN Ia rate (see \cite{wickramasinghe2000a}).

A catalog of 112 massive (with masses $\ge 0.8 M_{\odot}$) WDs
(both single and binary) that do not interact with other stars is
presented in \cite{nalezyty2004a}. The mass distribution for these
dwarfs is generally continuous, and decrease fairly rapidly with
increasing mass. Massive WDs with strong magnetic fields display
an almost flat mass distribution. Note that the four most massive
WDs (with masses $\ge 1.3 M_{\odot}$)are magnetic. The presence of
a magnetic field is not correlated with the temperature (and
probably also not with the age) of the white dwarf: magnetic and
non-magnetic degenerate dwarfs have the same temperature
distributions \cite{nalezyty2004a}. Table 1 in
\cite{nalezyty2004a}, shows that 3 of 25 MWDs are components of
binaries. Only one of these three dwarfs (EUVE J0317-855) is among
the most massive; it is situated in a very wide system, with an
approximate component separation of 1000 AU(see, for example,
\cite{ferrario1997a}); consequently, these stars have never
interacted. The high rotational velocity, high mass, and an age
inconsistent with the suggested companion (the WD LB 9802) suggest
that EUVE J0317-855 may be the result of a merger of two WDs
\cite{wickramasinghe2000a,vennes1996a,vennes1997a}.

Mergers of rotating, close, binary, stellar-mass black holes yield
peculiar consequences \cite{lousto2008a,komossa2008a}. Modeling
has shown that the black holes formed as a result of the merger
can acquire velocities of several thousands of km/s. These stars
subsequently leave not only their parent galaxies, but also the
clusters of galaxies to which they belong. Such black holes can
have covered distances of $\approx 50$ Mpc over the Hubble time.
Thus, they form a kind of continuous medium filling the space
between clusters of galaxies located at distances of up to 35-70
Mpc \cite{tikhonov2007a}. We have calculated the rates of mergers
of neutron stars (NSs) with WDs and with black holes (BHs). Our
aim was to obtain theoretical merger rates for various compact
objects (WDs, NSs, BHs), using the same computer code (the
``Scenario Machine'') to carry out population syntheses for the
same sets of parameters of the evolutionary scenario. A thorough
study of mergers of NSs with NSs and BHs and of the impact of
various parameters of the evolutionary scenario on these events
are presented, for example, in \cite{lipunov1996a,lipunov1997a}.

The current state of studies of the evolutionary status of Ap and
Bp stars is described in detail in \cite{north2008a}. We are
primarily interested in the occurrence rate for magnetic Ap stars
among all A stars. Roughly, $R=\frac{N(Ap)}{N(Ap)+N(A)}$ (here,
N(Ap) is the number of magnetic Ap stars and N(A) the number of A
stars). If we restrict the volume in which this relation is
measured, R becomes equal to 0.01-0.02 \cite{power2008a}. Here, we
suggest that magnetic Ap stars are the products of mergers of
components in close binaries with convective envelopes due to the
action of the magnetic stellar wind (MSW)\footnote{We will
consider the magnetic stellar wind only from the primary (more
massive) component}. In this case, the maximum and minimum initial
semimajor axes in such systems are $a_{max}/R_{\odot}\simeq 7.25
M/M_{\odot}$ and $a_{min}/R_{\odot}\simeq 6 (M/M_{\odot})^{1/3}$
(ñì. \cite{tutukov1988a}), where $M$ is the mass of each of the
binary components, assumed for this estimate to be equal. Only
components in close binaries for which the initial masses of each
component exceed $\approx 0.75 M_{\odot}$ (with a total mass of
$\approx 1.5 M_{\odot}$). can merge. The interval of semimajor
axes in the merging systems increases from zero for $M\approx 0.75
M_{\odot}$ to $7-11R_{\odot}$ for $M=1.5 M_{\odot}$. The latter
can be estimated using the following equation describing the loss
of orbital angular momentum of the system via the MSW (see Eq.
(34) in \cite{tutukov1988a}):

\begin{equation}
\label{mzv} \frac{d\ln J}{dt}=-10^{-14}\frac{R_2^4(M_1+M_2)^2
R_{\odot}}{\lambda^2 a^5 M_1 M_{\odot}}\mbox{s}^{-1},
\end{equation}

\noindent where $M_2$ and $R_2$ are the mass and radius of the
component with the MSW, $M_1$ is the mass of the secondary, $a$ is
the semimajor axis of the orbit, and $\lambda$ is the MSW
parameter (assumed to be unity). The above interval for the
semimajor axes corresponds to a rate of formation of Ap stars of
several per cent of the total number of stars with initial masses
exceeding $\sim M_{\odot}$. In this scenario, the strong magnetic
fields of the progenitors of Ap stars explain the presence of
strong fields in Ap stars. The equatorial rotational velocities of
magnetic Ap stars for $R\simeq 2 R_{\odot}$ and $P_{orb}=0.5-5$
day are $20-200$ km/s; i.e., they can be high. Note also that the
low rotational velocity displayed by many Ap stars does not
unambiguously argue against the possible formation of Ap stars as
a result of mergers of main-sequence stars with convective
envelopes, since angular momentum could be lost together with some
of the matter in the course of the merger.

\section{Population synthesis}

The principles of the ``Scenario Machine'' have been described
repeatedly, and we will restrict our consideration to the
selection of the free parameters of the evolutionary scenario. A
detailed description of the code is presented in
\cite{lipunov1996a,Lipunov1996,scm2}.

For each set of evolutionary-scenario parameters, we carried out a
population synthesis for $10^6$ binaries. The rates of events and
the numbers of objects in the Galaxy are obtained, assuming that
all stars are binary. In our study, we took as free parameters the
efficiency of the common envelope stage, $\alpha_{CE}$, and the
additional kick velocity acquired during the formation of neutron
stars and black holes.

\subsection{Kick Acquired in a Supernova Explosion}

We assumed that the NS or BH that forms during a supernova
explosion can acquire a kick velocity $v$, which is random and has
a Maxwell distribution:

\begin{equation}
\label{Maxwell} f(v)\sim \frac{v^2}{v^3_0}e^{-\frac{v^2}{v_0^2}},
\end{equation}

\noindent with all directions for the kick being equally probable.
The characteristic velocity $v_0$, acquired by the remnant is one
of the critical parameters for estimating the birth rates for
systems that remain bound after the supernova explosions in them.
Naturally, the results of the population synthesis strongly depend
on $v_0$. Increasing $v_0$ to velocities above the orbital
velocities in close binaries ($\sim 100$ km/s) results in a
dramatic decrease in the number of systems containing relativistic
components \cite{lipunov1997a}. Note, however, that this kick can
not only disrupt, but also bind some systems that would have
decayed without it.

It is believed that the magnitude of the kick velocity acquired
during the formation of a BH depends on the fraction of mass lost
by the star in the supernova explosion. We assume that a star
loses half its mass in the explosion \cite{bogomazov2005}. Thus,
for a BH, we take the characteristic kick velocity $v_0$ and
obtain the velocity $v$, according to the distribution (2), which
we divide by two.

\subsection{Mass-loss Efficiency in the Common-Envelope Stage}

During the common-envelope stage, stars transfer angular momentum
to the adjacent matter of the envelope very efficiently in course
of spiral-in. The efficiency of the mass loss in the
common-envelope stage is described by the parameter
$\alpha_{CE}=\Delta E_b/\Delta E_{orb}$, where $\Delta
E_b=E_{grav}-E_{thermal}$ is the binding energy of the ejected
envelope matter and $\Delta E_{orb}$ is the decrease in the
orbital energy of the system as the stars approach
\cite{tutukov1988a,heuvel1994a}. Thus, we obtain

\begin{equation}
\alpha_{CE} \left( \frac{GM_a M_c}{2a_f} - \frac{GM_a M_d}{2a_i}
\right)=\frac{GM_d(M_d - M_c)}{R_d},
\end{equation}

\noindent where $M_c$ is the core mass of the mass-losing star,
which has an initial mass $M_d$ and radius $R_d$ (which is a
function of the initial semimajor axis $a_i$ and the initial mass
ratio $M_a/M_d$, where $M_a$ is the mass of the accretor).

\subsection{Other Parameters of the Evolutionary Scenario}

In this Section, we present values for some parameters of the
evolutionary scenario. The maximum mass of a NS that can be
reached in the course of accretion (the Oppenheimer-Volkov limit),
was taken to be $M_{OV}=2.0 M_{\odot}$, while the initial masses
of young NSs were randomly distributed in the interval $1.25-1.44
M_{\odot}$ We assumed that main sequence stars with initial masses
in the interval $10-25 M_{\odot}$ end their evolution as NSs. We
augmented the progenitors of NSs with main sequence components of
close binaries that increase their mass as a result of mass
transfer, after which their mass reaches the above interval. More
massive stars ultimately evolve into BHs, and less massive stars
into WDs. In the calculations carried out here, we adopted an
equiprobable (flat) distribution for the component mass ratios of
the initial binaries \cite{tutukov1988a} and a zero initial
eccentricity for their orbits. All the calculations presented here
assume a classical weak stellar wind (see \cite{Lipunov1996}, and
also the A evolutionary scenario in \cite{bogomazov2005}). We also
adopted a flat distribution for the initial semimajor axes in the
binaries $d(\log a)=\mbox{const}$ in the interval $5-10^6
R_{\odot}$.

\section{Results}

Table 1 presents the birth and merger rates for WDs with various
chemical compositions, as well as the rates for WD mergers with
NSs and BHs, calculated for various mass loss efficiencies in the
common-envelope stage $\alpha_{CE}$. Table 2 contains the rates of
mergers of NSs with NSs and BHs for various characteristic kick
velocities acquired during the formation of the NSs ($v_0$) and
BHs ($v_0^{bh}$). The rates of events in Tables 1 and 2 are
calculated for a galaxy with a total mass of stars and gas of
$10^{11} M_{\odot}$ and a star formation rate specified by the
Salpeter function.

Figures 1a-d present the WD mass distributions. The solid curve
presents the distributions for WDs that have not undergone a
merger, and the dash-dotted curve the total mass distribution for
merging WDs. Figure 1a shows the distributions for WDs of all
chemical compositions, Fig. 1b for helium WDs, Fig. 1c for
carbon-oxygen WDs, and Fig. 1d for oxygen-neon WDs. Figure 1e
plots the total mass distribution for merging WDs with various
chemical compositions: the solid curve indicates mergers of
oxygen-neon WDs with carbon-oxygen WDs, the dash-dotted curve
mergers of oxygen-neon WDs with helium WDs, and the dotted curve
mergers of carbon-oxygen WDs with helium WDs.

Figures 2a-c present the average mass of merging WDs as a function
of the time since the onset of the formation of these stars.
Depending on the mass of the formed binary WD and the semimajor
axis of the WD binary, the time until the merger due to the action
of gravitational waves can range from several tens of million
years up to the Hubble time. The number of mergers of WDs
$N_\delta(t)$ was calculated assuming all stars in a galaxy are
born simultaneously (with the star formation rate represented by a
$\delta$ function). The evolution of this quantity in a galaxy
with arbitrary star formation is specified by the formula

\begin{equation}
 N(t)=\int\limits^{t}_{0} N_\delta(t-\tau)
 \varphi(\tau)\textit{d}\tau\label{not},
\end{equation}

\noindent where $N(t)$ is the number of mergers as a function of
time $t$ since the onset of star formation and $\varphi(t)$ a
function describing the star-formation history. The average mass
of merging dwarfs $\overline{M}(t, t+\Delta t)$ is given by

\begin{equation}
\overline{M}(t, t+\Delta t)=\frac{\sum\limits_i M_i}{N(t, t+\Delta
t)}, \label{mot}
\end{equation}

\noindent here $M_i$ are sums of the masses of WDs merging at a
time in the interval $t$ to $t+\Delta t$. Figure 2a presents the
average masses of merging WDs calculated assuming that all stars
in the galaxy form simultaneously (the function $N_\delta(t)$,
calculated using the ``Scenario Machine''); a constant star
formation rate was assumed in Fig. 2b; the star formation history
presented in \cite{kurbatov2005a} was used to construct the curves
in Fig. 2c; the numbers of mergers for the graphs in Fig. 2 band
2c were calculated using (4). The numbers in Figs 2a-2c denote the
average masses of 1 all merging WDs, 2 merging WDs with all
possible chemical compositions (but including only mergers in
which the total mass of the WDs exceeds the Chandrasekhar limit),
3 carbon–oxygen WDs, 4 carbon-oxygen WDs whose total mass exceeds
the Chandrasekhar limit.

The time since the formation of a binary WD system that can later
merge due to the radiation of gravitational waves depends on the
masses of the WDs in the system, as well as the semimajor axis of
the system. Figure 3 presents the dependence of the number of
mergers of carbon-oxygen WDs whose total masses exceed the
Chandrasekhar limit on the time since the formation of the close
binary containing two CO dwarfs.

Figure 4 presents the mass distribution for Ap stars, here
considered to be main sequence stars formed in mergers of main
sequence stars with convective envelopes under the action of the
magnetic stellar wind. Most of these have masses in the interval
$1.7-2.5 M_{\odot}$.

\section{Conclusion}

We can see from Fig. 1a that the number of WDs with low masses
($\lesssim 1 M_{\odot}$) originating due to mergers\footnote{With
the total mass of the merging dwarfs below the Chandrasekhar
limit.} is substantially smaller than the number of WDs that have
never undergone merging. The mass distribution for WDs that are
the end products of stellar evolution rather than of mergers of
other WDs rapidly decreases towards higher mass, while the mass
distribution for merging WDs with masses exceeding the solar value
is essentially flat. The numbers of WDs originating in mergers and
WDs that have not undergone merging become roughly equal for
masses near $1.1 M_{\odot}$. Naturally, at the highest masses
($\approx 1.3 M_{\odot}$), the number of WDs that form due to
merging exceeds the number that have never undergone merging by
roughly a factor of ten. This result is in good consistency with
the observations: on average, MWDs are more massive than WDs in
general, and the number of magnetic stars among the most massive
dwarfs is an order of magnitude higher than the number of
non-magnetic stars \cite{valyavin1999a}. When comparing Figs. 2
and 3 in \cite{nalezyty2004a}, we see also that MWDs dominate in
the mass distributions for magnetic and non-magnetic WDs with
masses $\approx 1.2-1.3M_{\odot}$ though the statistics are fairly
poor. Figures 1d and 1e show that the main contribution to the
total mass distribution in the interval $1.1-1.4M_{\odot}$ is made
by mergers of carbon-oxygen WDs, with a small contribution also
made by mergers of oxygen-neon with carbon-oxygen WDs. Thus, we
conclude that nearly all of the most massive MWDs originated as a
result of mergers of close binary degenerate dwarfs.

Figures 2a-2c show that the average mass of merging WDs (without
dividing them into different types according to their chemical
composition) decreases rapidly with the age of the population,
from $\approx 2.1 M_{\odot}$ to less than $1 M_{\odot}$. However,
the average mass of merging WDs whose total mass exceeds the
Chandrasekhar limit decreases rapidly from the maximum value to
($\approx 1.8-1.9 M_{\odot}$), after which it remains nearly
constant with time. The same pattern is observed if we restrict
our consideration to the most popular candidate SN Ia progenitors
-- mergers of carbon-oxygen dwarfs with carbon-oxygen dwarfs. In a
young galaxy, the average total mass of merging CO dwarfs is about
$1.9 M_{\odot}$. For times $\lesssim 1\cdot 10^9$ years, this
decreases to $1.6-1.7 M_{\odot}$ then remains the same for the
subsequent 13 billion years of evolution. Thus, we conclude that
SN Ia can be used as standard candles only in galaxies older than
approximately one billion years. If we assume that star formation
begins at redshift\footnote{Here we use the simple formula
$t(z)=\frac{2}{3H_0 (1+z)^{3/2}}$, where $t$ is the time since the
Big Bang, $z$ the redshift, and $H_0$ the Hubble constant, which
in our approximate calculations is taken to be 75 km s$^{-1}$
Mpc$^{-1}$.} z=10, then the average total mass of merging CO
dwarfs will stop decreasing only at redshifts z îêîëî of about two
to three. This conclusion does not depend on our selection of
star-formation functions (from those considered in the present
study). Figure 1c indicates that the maximum mass of merging CO
dwarfs is roughly $2.1 M_{\odot}$, which exceeds the Chandrasekhar
limit by a factor of 1.5. If we suppose that the energy release at
the time of the thermonuclear explosion of the merged dwarfs is
proportional to the mass, we expect that the scatter in the
brightnesses of SN Ia at their maximum luminosity is also roughly
a factor of 1.5 (or $0^m.5$). The conclusion that expansion of the
Universe was accerating reported in
\cite{riess1998a,perlmutter1999a} (Figures 4 and 5 in
\cite{riess1998a} and Figure 2 in \cite{perlmutter1999a}) was
based on observations of SN Ia at distances of up to $z\sim 1$.
The errors in the supernova brightnesses in these studies are
about $0^m.5$. Consequently, the evolution of the average total
mass of merging WDs and the scatter of the total mass about the
average could not substantially affect the main conclusions of
\cite{riess1998a,perlmutter1999a}. However, it is essential to
take into account described decrease of the total mass of merging
white dwarfs with time in accurate estimations of parameters of
acceleration of the Universe. Chemical composition variations may
introduce additional scatter into the brightnesses of distant,
``young'' SN Ia; however, the present estimates of this effect
suggest that this factor can be neglected \cite{meng2008a}.

Figure 3 indicates that the minimum time for two carbon-oxygen WDs
whose total mass exceeds the Chandrasekhar limit to merge is
roughly 90 thousand years. During the first million years of their
lives, roughly 1\% of such binary CO dwarfs merge. Consequently,
in most cases, the envelopes of SN Ia (considered in
\cite{patat2008a}) cannot consist of matter ejected by the stars
during the common-envelope stage, and the formation of SN Ia via
the merging of two CO dwarfs in systems with envelopes is
unlikely.

According to our calculations, the rate of mergers of main
sequence stars under the action of the magnetic stellar wind for
masses from $0.75 M_{\odot}$ to $1.5 M_{\odot}$ is roughly
0.0037/year in the Galaxy, while the total number of main-sequence
stars in the Galaxy formed due to mergers for masses from 1.5 to
$3 M_{\odot}$ is roughly $2.7\cdot 10^6$. The total birth rate for
stars with masses of 1.5-$3M_{\odot}$ is 0.26/year, with the total
number of such stars in the Galaxy being $2.4\cdot 10^8$. Thus,
the fraction of main-sequence stars formed due to mergers of
components due to the action of the MSW in the indicated mass
interval given by the population synthesis computations is in good
agreement with the fraction of Ap stars among A-type stars in a
restricted volume \cite{power2008a}. The mass distribution for
stars formed by mergers differs from a Salpeter distribution (Fig.
4). An attempt to distinguish the mass and age distributions for
ordinary A and B stars and magnetic Ap and Bp stars was made, for
example, in \cite{hubrig2000a}, but the available statistics were
too poor to enable unambiguous conclusions about whether the mass
and age distributions for magnetic and ordinary A and B stars are
similar or different \cite{north2008a}. Therefore, based on the
results of our population synthesis, we conclude that at least
some magnetic Ap and Bp stars could be formed by mergers of main
sequence stars with convective envelopes under the action of the
MSW\footnote{4We assumed that the magnetic stellar wind alters the
rotational angular momentum in accordance with (1) in systems
containing two main sequence stars with convective envelopes, when
the masses of both stars are within $0.75 M_{\odot}$--$1.5
M_{\odot}$. We also assumed that the matter of the components
mixes totally after the merger. }; however, this hypothesis still
awaits observational confirmation.

\section{Acknowledgments}

The population syntheses for close binaries for this study were
carried using the ``Scenario Machine'' code developed by V.M.
Lipunov, K.A. Postnov, and M.E. Prokhorov \cite{Lipunov1996} in
the Department for Relativistic Astrophysics of the Sternberg
Astronomical Institute, Moscow State University.

The work was supported by the State Program of Support for Leading
Scientific Schools of the Russian Federation (grant
NSh-1685.2008.2, AIB; grant NSh-4354.2008.2, AVT), the Analytical
Departmental Targeted Program ``The Development of Higher
Education Science Potential'' (grant RNP-2.1.1.5940, AIB), and the
Russian Foundation for Basic Research (project code 08-02-00371,
AVT).

\subsection*{Figure captions}

\begin{figure*}[h!]
\hspace{0cm} \epsfxsize=0.4\textwidth\centering \epsfbox{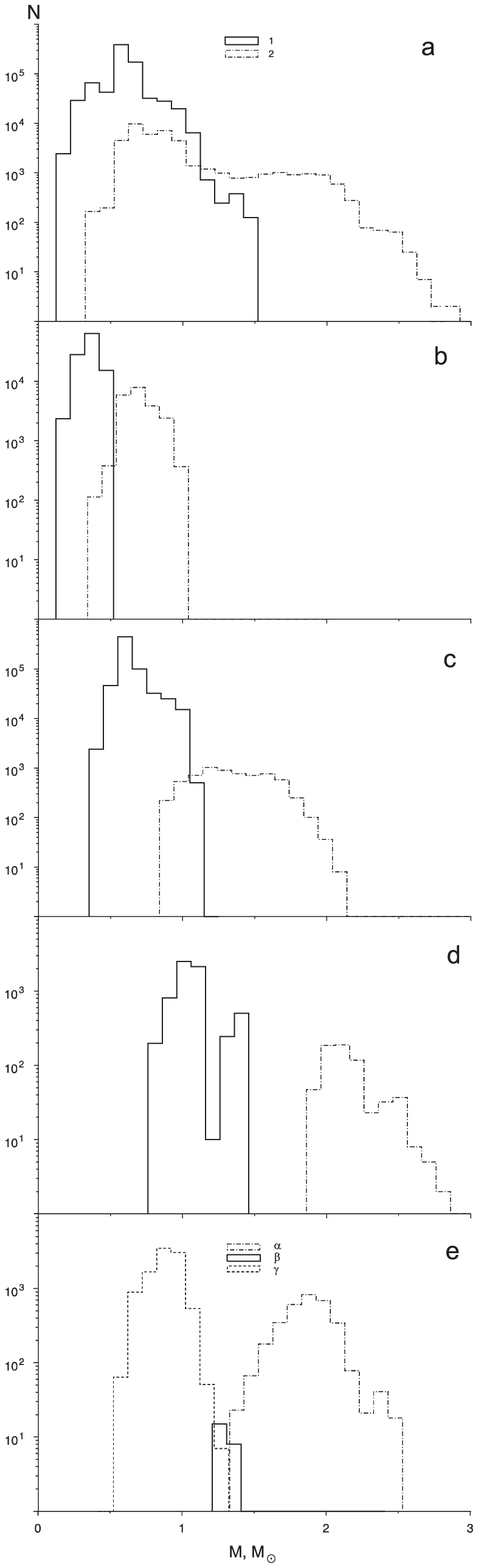}
\vspace{0cm}\caption{} \label{fig1}
\end{figure*}

Figure \ref{fig1}: The mass distribution for WDs that have not
undergone mergers (solid curve) and the total mass distribution
for merging WDs (dash-dotted curve). The horizontal axis plots the
mass of the WD in $M_{\odot}$ (the total mass of the two WDs in
the case of merging WDs), and the vertical axis the number of WDs
(and of mergers) derived from the calculations. Graph (a)
corresponds to all types of chemical compositions of WDs, (b) to
helium WDs, (c) to carbon-oxygen WDs, (d) to oxygen-neon WDs, and
graph (e) presents the total mass distribution for merging WDs of
various chemical compositions. In graph (e), $\alpha$ denotes
ONe+CO, $\beta$ ONe+He, and $\gamma$ CO+He.

\bigskip

\begin{figure*}[h!]
\hspace{0cm} \epsfxsize=0.6\textwidth\centering \epsfbox{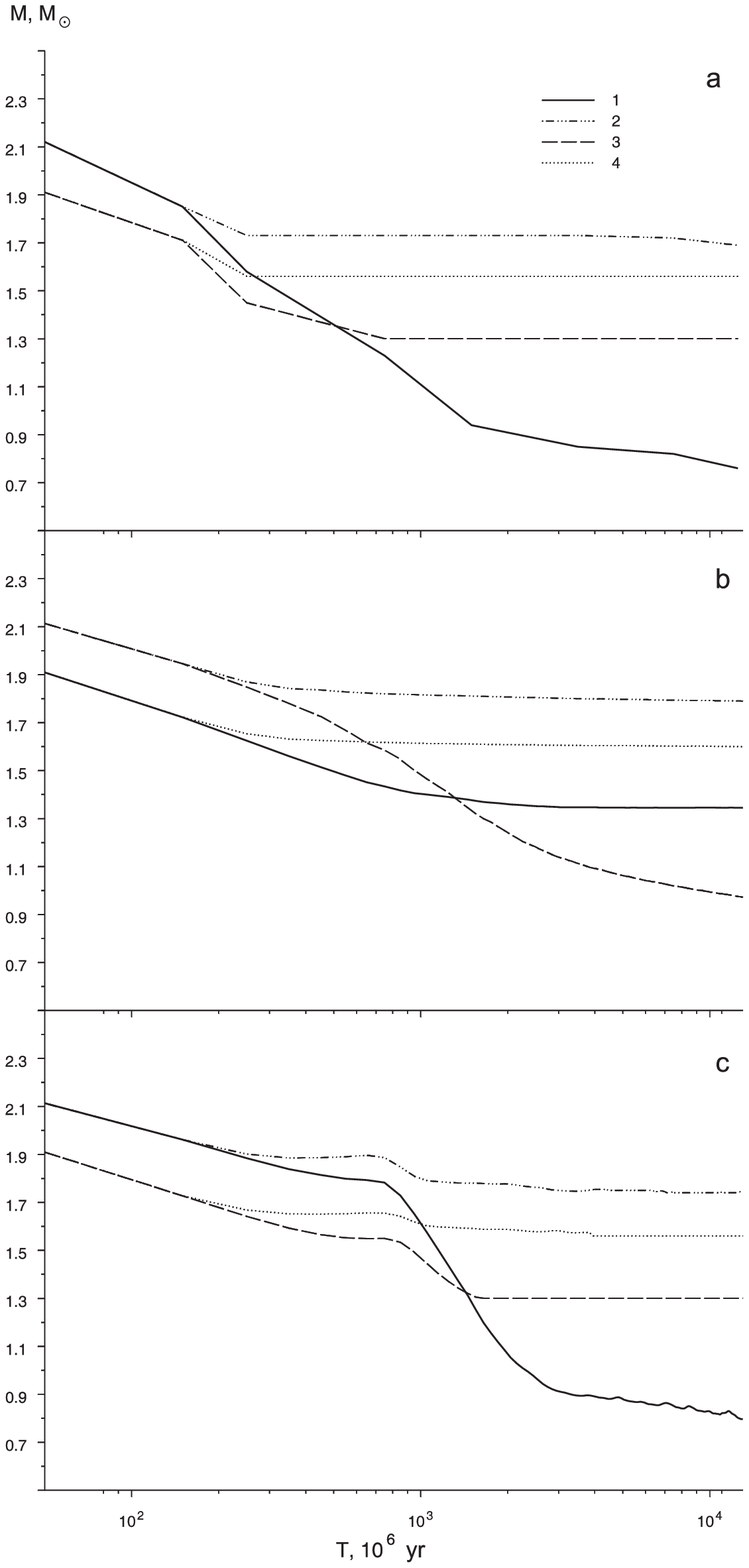}
\vspace{0cm}\caption{} \label{fig2}
\end{figure*}

Figure \ref{fig2}: The average mass of merging WDs as a function
of time from the onset of the star formation. The numbers at the
top mark the curves corresponding to mergers of 1 all WDs, 2 WDs
of all possible chemical compositions, but taking into account
only mergers in which the total mass of the WDs exceeds the
Chandrasekhar limit, 3 carbon-oxygen WDs, and 4 carbon-oxygen WDs
with the total mass exceeding the Chandrasekhar limit. The
following assumptions for the star formation history were made:
(a) all stars in a galaxy were formed at the same initial time,
(b) the star formation rate has been constant, (c) the star
formation history corresponds to that in \cite{kurbatov2005a}.

\bigskip

\begin{figure*}[h!]
\hspace{0cm} \epsfxsize=0.6\textwidth\centering \epsfbox{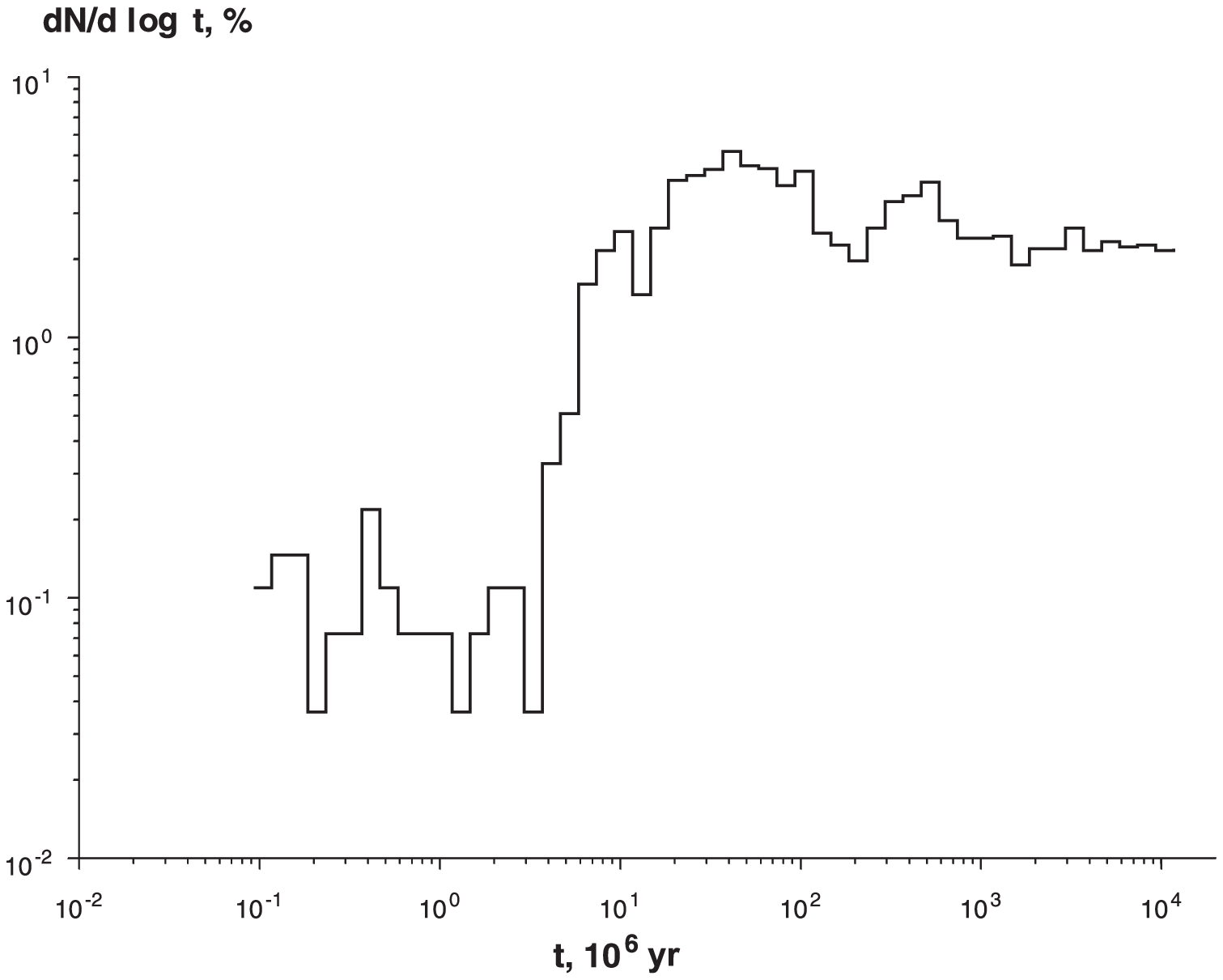}
\vspace{0cm}\caption{} \label{fig3}
\end{figure*}

Figure \ref{fig3}: Distribution of the lifetime from the formation
of a binary system consisting of two CO dwarfs to their merger (SN
Ia explosion), for carbon-oxygen WD binaries with total masses
exceeding the Chandrasekhar limit.

\bigskip

\begin{figure*}[h!]
\hspace{0cm} \epsfxsize=0.6\textwidth\centering \epsfbox{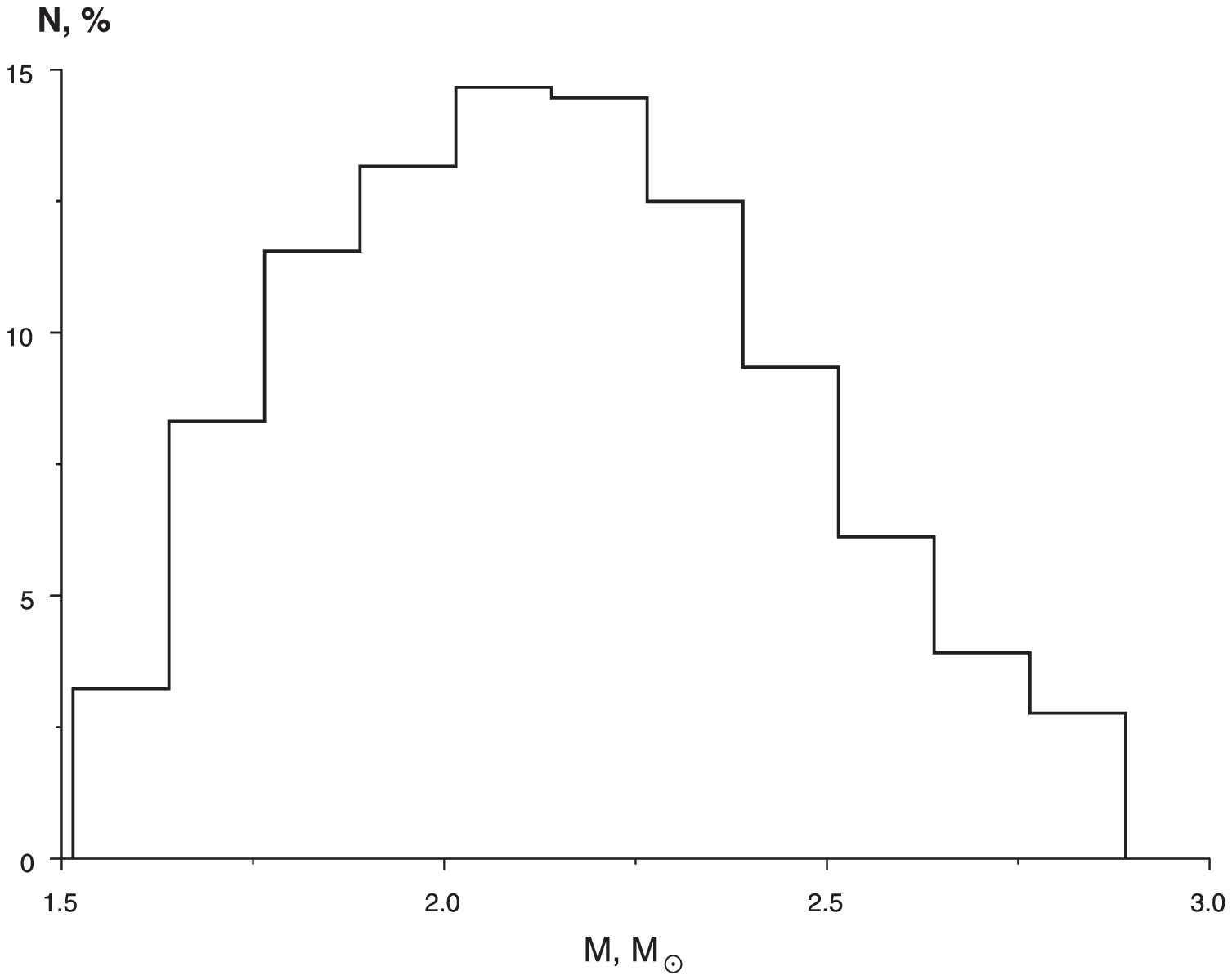}
\vspace{0cm}\caption{} \label{fig4}
\end{figure*}

Figure \ref{fig4}: Mass distribution for main sequence stars
formed in mergers of main sequence stars with convective envelopes
under the action of the magnetic stellar wind.

\newpage

\begin{table}
\caption{ Birth and merger rates for various types of WD, and
rates of mergers of WDs with NSs and BHs, calculated for various
efficiencies of the common envelope stage $\alpha_{CE}$, per year
in the Galaxy. }\label{wdfreq}
\newcolumntype{C}{>{\centering\arraybackslash}m}
\newcolumntype{L}{>{\arraybackslash}m}
\begin{tabular}{|L{30mm}|C{25mm}|C{25mm}|C{25mm}|}
\hline
Event $\backslash$ $\alpha_{CE}$ & 0.3 & 0.5 & 1.0 \\
\hline
He+He$^1$ & 0.016 & 0.015 &  0.011 \\
He+Any$^2$ & 0.066 & 0.08 & 0.1 \\
CO+CO$^1$ & $4.7\cdot 10^{-3}$ & $4.8\cdot 10^{-3}$ & $4.4\cdot 10^{-3}$ \\
CO+CO$^{1,3}$ & $1.9\cdot 10^{-3}$ & $1.9\cdot 10^{-3}$ & $1.8\cdot 10^{-3}$ \\
CO+Any$^2$ & 0.49 & 0.49 & 0.5 \\
ONe+ONe$^1$ & $1.1\cdot 10^{-4}$ & $4.7\cdot 10^{-4}$ & $7.7\cdot 10^{-4}$ \\
ONe+Any$^2$ & 0.004 & $4.7\cdot 10^{-3}$ & $5.4\cdot 10^{-3}$ \\
ONe+He$^1$ & $3.5\cdot 10^{-5}$ & $6\cdot 10^{-5}$ & $2.8\cdot 10^{-4}$ \\
CO+He$^1$ & 0.006 & $8.5\cdot 10^{-3}$ & $8.2\cdot 10^{-3}$ \\
ONe+CO$^1$ & 0.002 & $2.4\cdot 10^{-3}$ & $1.8\cdot 10^{-3}$ \\
NS+WD$^1$ & $4.7\cdot 10^{-4}$ & $4.7\cdot 10^{-4}$ & $6.1\cdot 10^{-4}$ \\
BH+WD$^1$ & 0 & $1.5\cdot 10^{-6}$ & $3.7\cdot 10^{-6}$ \\
\hline \multicolumn{4}{|l|}{\text{\footnotesize $^1$Merging of corresponding objects.}} \\
\multicolumn{4}{|l|}{\text{\footnotesize $^2$Formation of corresponding systems.}} \\
\multicolumn{4}{|l|}{\text{\footnotesize $^3$The total mass of the merging dwarfs exceeds the Chandrasekhar limit.}} \\
\hline
\end{tabular}
\end{table}

\begin{table}
\caption{ Rates for mergers of NSs with NSs and BHs calculated for
varioius kick velocities obtained during the formation of the NSs
($v_0$) and ($v_0^{bh}$), per year in the Galaxy.}\label{nsbhfreq}
\newcolumntype{C}{>{\centering\arraybackslash}m}
\begin{tabular}{|C{35mm}|C{20mm}|C{20mm}|C{20mm}|}
\hline
Event & \multicolumn{3}{|c|}{\text{$v_0$, $v_0^{bh}$, km/s}} \\
\hhline{|~|-|-|-|}
& 0, 0 & 100, 0 & 100, 100 \\
\hline
NS+NS & $5.5\cdot 10^{-4}$ & $1.3\cdot 10^{-4}$ & $1.3\cdot 10^{-4}$ \\
BH+NS & $1.5\cdot 10^{-4}$ & $1.6\cdot 10^{-4}$ & $1.3\cdot 10^{-4}$ \\
BH+BH & $9.8\cdot 10^{-5}$ & $9.8\cdot 10^{-5}$ & $1.2\cdot 10^{-4}$ \\
\hline \multicolumn{4}{|l|}{\text{\footnotesize The velocity of the BH was taken from (2) with $v_0^{bh}$, then halved.}} \\
\hline
\end{tabular}
\end{table}

\begin{thebibliography}{99}\addcontentsline{toc}{chapter}{References}

\bibitem{tutukov1981a} A. V. Tutukov and L. R. Yungel'son, Nauchn. Inform.
Astron. Sovet Akad. Nauk SSSR {\bf 49}, 3 (1981).

\bibitem{iben1984a} I. Jr. Iben, A. V. Tutukov, Astrophys. J. Suppl. Ser. {\bf 54}, 335
(1984).

\bibitem{li2001a} W. Li, A. V. Filippenko, R. R. Treffers, A. G. Riess, J. Hu, Y. Qiu, Astrophys. J. {\bf 546},
734 (2001).

\bibitem{whelan1973a} J. Whelan, I. J. Iben, Astrophys. J. {\bf 186}, 1007
(1973).

\bibitem{livne1990a} E. Livne, Astrophys. J. {\bf 354}, L53 (1990).

\bibitem{aubourg2007a} E. Aubourg, R. Tojeiro, R. Jimenez, et al., eprint
arXiv:0707.1328 (2007).

\bibitem{yoon2007a} S.-C. Yoon, Ph. Podsiadlowski, S. Rosswog, Mon. Not. Roy. Astron. Soc. {\bf 380}, 933 (2007).

\bibitem{riess1998a} A. G. Riess, A. V. Filippenko, P. Challis, et
al., Astron. J. {\bf 116}, 1009 (1998).

\bibitem{perlmutter1999a} S. Perlmutter, G. Aldering, G. Goldhaber, et al., Astrophys. J. {\bf 517}, 565
(1999).

\bibitem{pskovskii1984a} Yu. P. Pskovsky, Sov. Astron. {\bf 28},
658 (1984).

\bibitem{taubenberger2008a} S. Taubenberger, S. Hachinger, G. Pignata, et
al., Mon. Not. Roy. Astron. Soc. {\bf 385}, 75 (2008).

\bibitem{pignata2008a} G. Pignata, S. Benetti, P. A. Mazzali, et
al., eprint arXiv:0805.1089, MNRAS, in press (2008).

\bibitem{roepke2006a} F. K. Roepke, Reviews in Modern Astronomy {\bf 19}, 127
(2006).

\bibitem{blanc2008a} G. Blanc, L. Greggio, New Astronomy {\bf 13}, 606 (2008).

\bibitem{shustov1997a} B. Shustov, D. Wiebe, A. Tutukov, Astron.
Astrophys. {\bf 317}, 397 (1997).

\bibitem{forster2008a} F. F\"orster, K. Schawinski, Mon. Not. Roy. Astron. Soc. Lett. {\bf 388}, L74 (2008).

\bibitem{patat2008a} F. Patat, P. Chandra, R. Chevalier, et al.,
Messenger {\bf 131}, 30 (2008).

\bibitem{tutukov1972a} A. V. Tutukov and L. R. Yungelson, Astrofizika {\bf
8}, 381 (1972).

\bibitem{wickramasinghe2000a} D. T. Wickramasinghe, L. Ferrario, Publ. Astron. Soc. Pacific {\bf 112}, 873
(2000).

\bibitem{nalezyty2004a} M. Nalezyty, J. Madej, Astron. Astrophys.
{\bf 420}, 507 (2004).

\bibitem{ferrario1997a} L. Ferrario, S. Vennes, D. T. Wickramasinghe, et al., Mon. Not. Roy.
Astron. Soc. \textbf{292}, 205 (1997).

\bibitem{vennes1996a} S. Vennes, P. A. Thejll, D. T. Wickramasinghe, M. S. Bessell, Astroph. J. {\bf 467},
782 (1996).

\bibitem{vennes1997a} S. Vennes, P. A. Thejll, R. G. Galvan, J. Dupuis, Astrophys. J. {\bf 480}, 714
(1997).

\bibitem{lousto2008a} C. Lousto, Y. Zlochower, eprint
arXiv:0805.0159 (2008).

\bibitem{komossa2008a} S. Komossa, H. Zhou, H. Lu, Astrophys. J.
{\bf 678}, L81 (2008).

\bibitem{tikhonov2007a} A. Tikhonov, Astronomy Letters {\bf 33},
499 (2007).

\bibitem{lipunov1996a} V. M. Lipunov, K. A. Postnov, M. E. Prokhorov, Astron. Astrophys. {\bf 310},
489 (1996).

\bibitem{lipunov1997a} V. M. Lipunov, K. A. Postnov, M. E. Prokhorov, Mon. Not. Roy. Astron. Soc. {\bf 288},
245 (1997).

\bibitem{north2008a} P. North, J. Babel, D. Erspamer, Contrib. Astron. Obs. Skalnate Pleso \textbf{38}, no. 2, 375
(2008).

\bibitem{power2008a} J. Power, G. Wade, M. Aurie`re, J. Silvester, D. Hanes, Contrib. Astron. Obs. Skalnate Pleso {\bf 38}, no. 2,
443 (2008).

\bibitem{tutukov1988a} A. G. Masevich and A. V. Tutukov, Star Evolution:
The Theory and Observations (Nauka, Moscow, 1988) [in Russian].

\bibitem{Lipunov1996} V. M. Lipunov, K. A. Postnov, M. E.
Prokhorov, Astrophysics and Space Physics Reviews {\bf 9}, 1
(1996, The Scenario Machine: Binary Star Population Synthesis,
Eds. R. A. Sunyaev, Harwood academic publishers).

\bibitem{scm2} V. M. Lipunov, K. A. Postnov, M. E. Prokhorov, A. I. Bogomazov, accepted by Astron. Rep., eprint arXiv:0704.1387 (2007).

\bibitem{bogomazov2005} A. I. Bogomazov, M. K. Abubekerov, and V. M. Lipunov, Astron. Rep. \textbf{49}, 644 (2005).

\bibitem{heuvel1994a} E. P. J. van den Heuvel, in S. N. Shore, M. Livio, E. P. J. van den Heuvel, Interacting Binaries, p.
103, Springer-Verlag, (1994).

\bibitem{kurbatov2005a} E. P. Kurbatov, A. V. Tutukov, and B. M. Shustov, Astron. Rep. {\bf 49},
510 (2005).

\bibitem{valyavin1999a} G. Valyavin, S. Fabrika, ASP Conference Series \textbf{169}, 206 (1999).

\bibitem{meng2008a} X. Meng, X. Chen, Z. Han, eprint
arXiv:0802.2471 (2008).

\bibitem{hubrig2000a} S. Hubrig, P. North, G. Mathys, Astrophys. J. {\bf 539}, 352 (2000).

\end{thebibliography}
\end{document}